\documentclass[
reprint,
superscriptaddress,
bibnotes,
amsmath,
amssymb,
aps,
prb,
floatfix,
]{revtex4-1}

\usepackage{graphicx}
\usepackage{dcolumn}
\usepackage{bm}
\usepackage{epstopdf}

\begin{document}

%Title of paper
\title{Neutron spin-echo study of the critical dynamics of spin-5/2 antiferromagnets in two and three dimensions}

\author{K. F. Tseng}
\affiliation{Max-Planck-Institut f\"ur Festk\"orperforschung, Heisenbergstrasse 1, D-70569 Stuttgart, Germany}
\affiliation{Max Planck Society Outstation at the Forschungsneutronenquelle Heinz Maier-Leibnitz (MLZ), D-85747 Garching, Germany}
\author{T. Keller}
\affiliation{Max-Planck-Institut f\"ur Festk\"orperforschung, Heisenbergstrasse 1, D-70569 Stuttgart, Germany}
\affiliation{Max Planck Society Outstation at the Forschungsneutronenquelle Heinz Maier-Leibnitz (MLZ), D-85747 Garching, Germany}
\author{A. C. Walters}
\affiliation{Max-Planck-Institut f\"ur Festk\"orperforschung, Heisenbergstrasse 1, D-70569 Stuttgart, Germany}
\affiliation{Diamond Light Source Limited, Chilton, Didcot, Oxfordshire OX11 0DE, United Kingdom} %\altaffiliation
\author{R. J. Birgeneau}
\affiliation{Materials Science Division, Lawrence Berkeley National Laboratory, Berkeley, CA 94720, USA}
\author{B. Keimer}
\email[To whom correspondence should be addressed. E-mail: ]{b.keimer@fkf.mpg.de}
\affiliation{Max-Planck-Institut f\"ur Festk\"orperforschung, Heisenbergstrasse 1, D-70569 Stuttgart, Germany}

%\date{\today}

\begin{abstract}
We report a neutron spin-echo study of the critical dynamics in the $S=5/2$ antiferromagnets MnF$_2$ and Rb$_2$MnF$_4$ with three-dimensional (3D) and two-dimensional (2D) spin systems, respectively, in zero external field. Both compounds are Heisenberg antiferromagnets with a small uniaxial anisotropy resulting from dipolar spin-spin interactions, which leads to a crossover in the critical dynamics close to the N\'eel temperature, $T_N$. By taking advantage of the $\mu\text{eV}$ energy resolution of the spin-echo spectrometer, we have determined the dynamical critical exponents $z$ for both longitudinal and transverse fluctuations. In MnF$_2$, both the characteristic temperature for crossover from 3D Heisenberg to 3D Ising behavior and the exponents $z$ in both regimes are consistent with predictions from the dynamical scaling theory. The amplitude ratio of longitudinal and transverse fluctuations also agrees with predictions. In Rb$_2$MnF$_4$, the critical dynamics crosses over from the expected 2D Heisenberg behavior for $T\gg T_N$ to a scaling regime with exponent $z = 1.387(4)$, which has not been predicted by theory and may indicate the influence of long-range dipolar interactions.
\end{abstract}

%\pacs{75.30.DS, 75.40.Gb, 75.50.Ee}

\maketitle

\section{Introduction}

Following the discovery of high-temperature superconductivity in doped antiferromagnets, the spin dynamics of both two-dimensional (2D) and three-dimensional (3D) antiferromagnets have received considerable attention in recent years. Since the spin systems of the parent compounds of the copper- and iron-based superconductors are nearly isotropic, \cite{keimer1992,greven1994,greven1995,kim2001} the spin excitations and critical dynamics of Heisenberg antiferromagnets have been widely studied by inelastic neutron scattering. \cite{birgeneau1990,nakajima1995,lee1998,leheny1999} The temperature dependence of the magnetic correlation lengths, $\xi$, in the paramagnetic state generally agree with scaling relations predicted by the theory of critical phenomena, \cite{collins1989,halperin1969,hohenberg1977} independent of whether the spins are in the classical or quantum limit. Because of the limited energy resolution of neutron triple-axis spectrometry (TAS), however, much less information is available on the energy widths, $\Gamma$, of the spin excitations in the paramagnetic state and their dynamical scaling behavior, $\Gamma \sim \xi^{-z}$, with the dynamical critical exponent $z$.

In RbMnF$_3$, one of the best experimental realizations of the three-dimensional Heisenberg antiferromagnet (3DHA), the dynamical critical exponent is in good agreement with the dynamical scaling theory which predicts $z = 1.5$. \cite{coldea1998} In MnF$_2$, where dipolar spin-spin interactions induce a small uniaxial anisotropy, the measured static exponents $\beta$, $\nu$, and $\gamma$ follow 3D Ising behavior, as expected close to the N\'eel temperature $T_N$, but the dynamic exponent $z$ is close to the value 1.5 predicted for the 3DHA. \cite{Heller1966,schulhof1971} This origin of this discrepancy has not yet been conclusively resolved, but it it is probably caused by the limited energy resolution of neutron three-axis spectroscopy (TAS), \cite{schulhof1971} with precludes inelastic scattering measurements sufficiently close to $T_N$.

The undoped parent compounds of the cuprate superconductors, such as La$_2$CuO$_4$, are excellent models for the two-dimensional Heisenberg antiferromagnet (2DHA) with $S=1/2$. The temperature dependent correlation length measured by neutron scattering is well described by theoretical work on the 2DHA, not only for $S=1/2$ compounds (Refs.~\onlinecite{keimer1992,greven1994,greven1995,kim2001}), but also for related compounds with $S=1$ (Refs.~\onlinecite{birgeneau1990,nakajima1995}) and $S=5/2$ (Refs.~\onlinecite{lee1998,leheny1999}). Measurements on the spin dynamics in the paramagnetic state of $S=1/2$ systems are in good agreement with the exponent $z =1$ predicted for the 2DHA. \cite{kim2001} For the quasi-2D $S=5/2$ compound Rb$_2$MnF$_4$, on the other hand, the uniaxial spin-space anisotropy is expected to generate a crossover from Heisenberg to Ising behavior upon cooling towards $T_N$, which precludes experimental tests of the dynamical scaling by TAS, as in the case of MnF$_2$. Neutron scattering data in a magnetic field $H$ close to the bicritical point in the $H-T$ phase diagram, where the anisotropy is expected to become irrelevant, yielded a value of $z=1.35 \pm 0.02$, clearly different from the theoretically predicted $z=1$. \cite{christianson2001} The origin of this unexpected exponent has thus far remained unresolved.

Motivated by these open questions, we have re-investigated the critical dynamics of the model compounds MnF$_2$ and Rb$_2$MnF$_4$ by means of the neutron spin-echo (NSE) triple-axis spectroscopy technique with energy resolution in the $\mu\text{eV}$ range. A related technique was first used by Mezei to study the critical dynamics of poly-crystalline iron~\cite{mezei1982,mezei1984} and later optimized for the measurement of linewidths of quasi-elastic excitations at small momentum transfer $\bm{Q}$. \cite{MezeiBible1980} For the present study at larger $\bm{Q}$,  we took advantage of a modified type of NSE based on radio-frequency spin flippers incorporated in a TAS spectrometer (termed neutron resonant spin-echo, NRSE). \cite{golub1987,gaehler1988} In this setup, the TAS provides good momentum resolution and helps suppress the background, but offers a comparatively coarse energy resolution, while the spin-echo device enhances the energy resolution by about two orders of magnitude. The neutron spin-flip processes related to the scattering by spin excitations lead to complicated spin-echo signals. To describe these effects, we introduce an analysis technique based on a ray-tracing simulation of the spectrometer. In this way, we are able to discriminate between longitudinal and transverse fluctuations at positions in $\bm{Q}$-space where both fluctuation components contribute to the scattering cross section. Since it has thus far proven difficult to find a scattering vector $\bm{Q}$ where only one of these components has a nonzero cross section, this is an additional distinct advantage of the NRSE-TAS setup.

\section{Experimental details}

\begin{figure}[t]
	\includegraphics[width=\columnwidth]{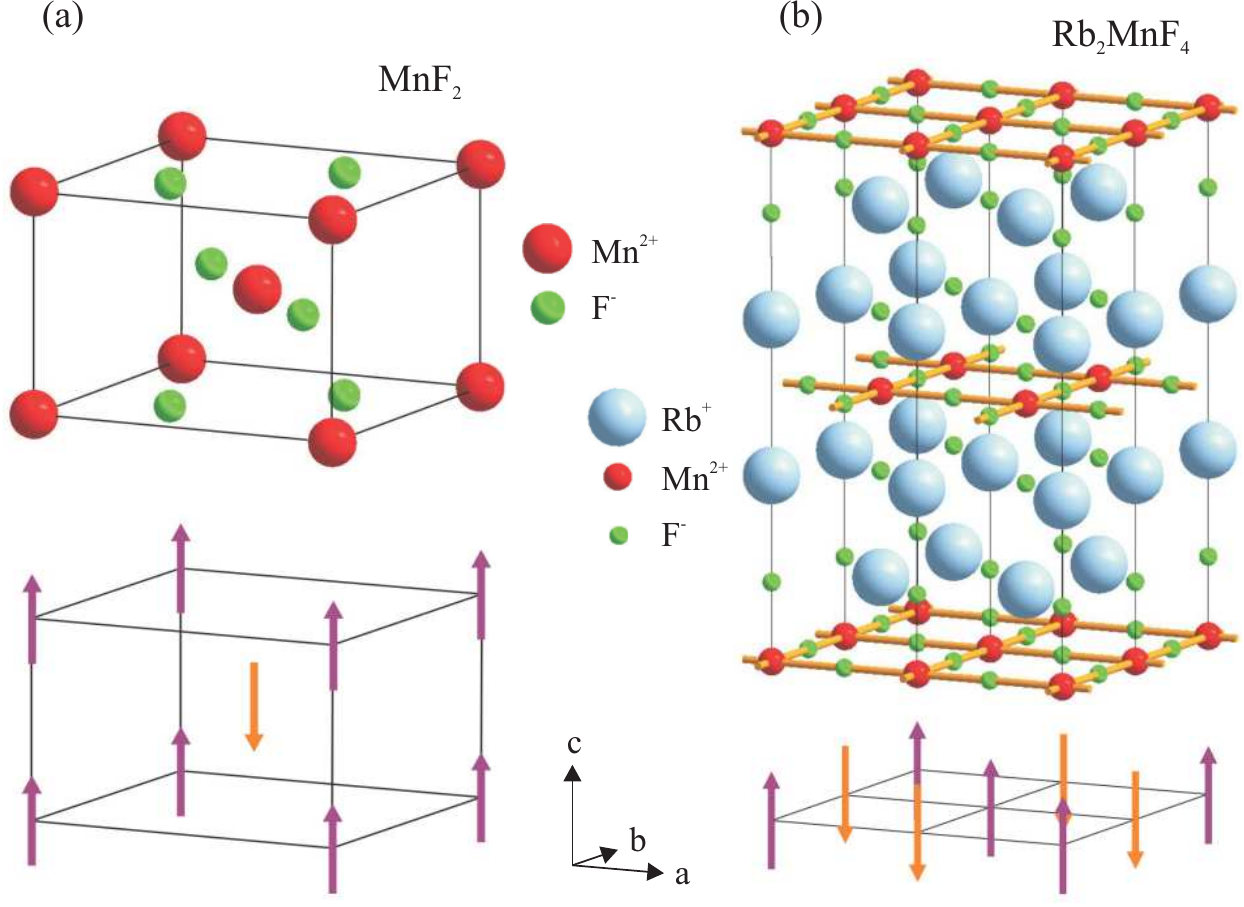}
	\caption{Chemical (top) and magnetic (bottom) structures of (a) MnF$_2$ and (b) Rb$_2$MnF$_4$. In the ordered state, the spins in both compounds are aligned along the tetragonal $c$-axis.}
\end{figure}

\begin{figure}[t]
	\includegraphics[width=\columnwidth]{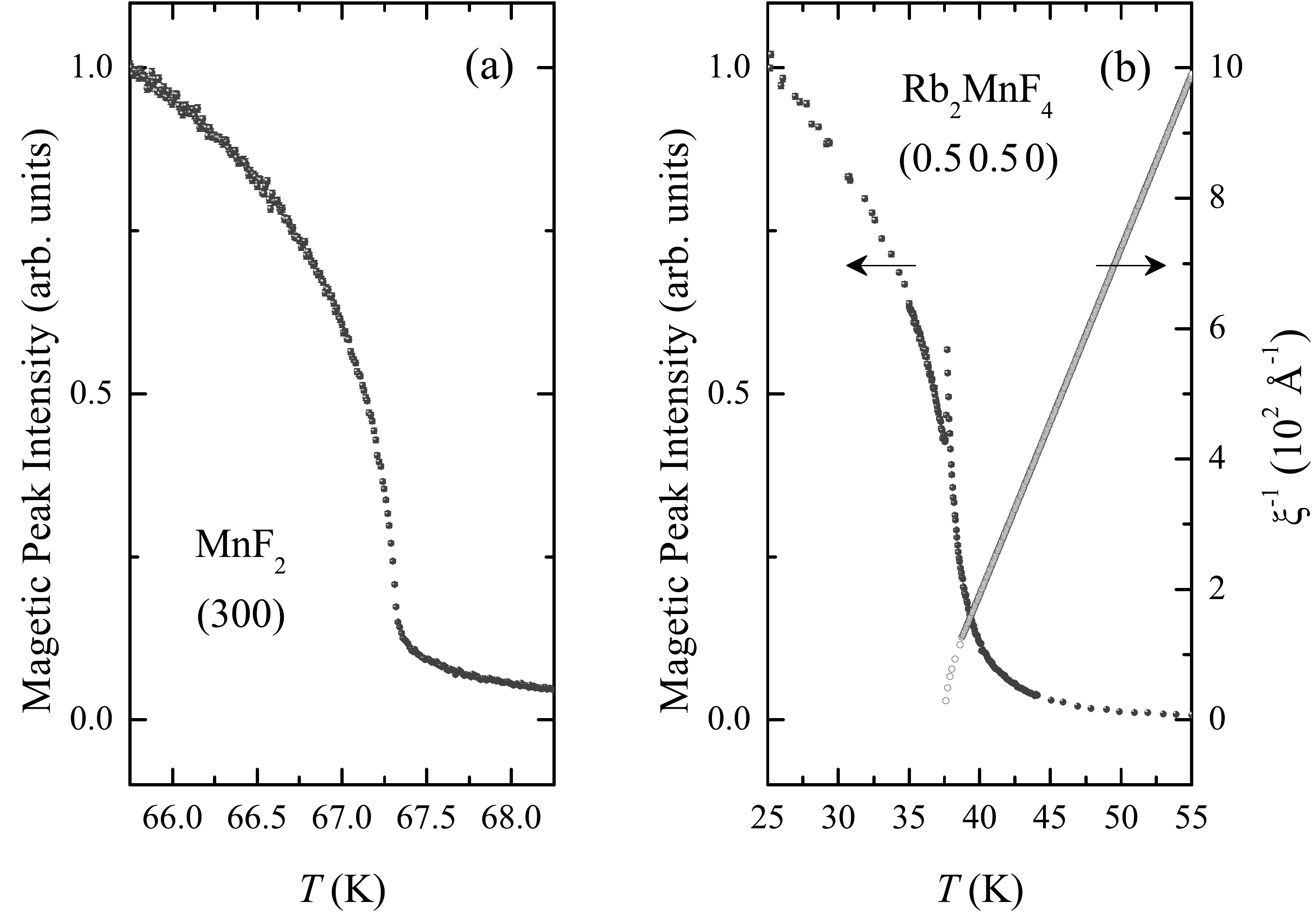}
	\caption{Antiferromagnetic order parameters. (a) Intensity of the antiferromagnetic Bragg peak $(300)$ in MnF$_2$ as a function of temperature. The maximum slope defines the N\'eel temperature $T_N$. (b) Left axis: Intensity of the $(0.5\,0.5\,0)$ magnetic Bragg reflection of Rb$_2$MnF$_4$. The sharp peak results from critical scattering and defines $T_N$. Right axis: Calculated inverse correlation length, $\xi^{-1}(T)$, for the 2D $S=5/2$ Heisenberg model with Ising anisotropy distortion according to Refs.~\onlinecite{cuccoli1996,cuccoli1997}.}
	\label{fig:orderparameter}
\end{figure}

MnF$_2$ and Rb$_2$MnF$_4$ are weakly anisotropic Heisenberg antiferromagnets with 3D and 2D spin systems, respectively. Both compounds form body-centered tetragonal crystal lattices. MnF$_2$ crystallizes in the rutile structure ($a= 4.874\,\text{\AA}$, $c=3.300\,\text{\AA}$), Rb$_2$MnF$_4$ in the K$_2$NiF$_4$ structure ($a= 4.230\,\text{\AA}$, $c= 13.82\,\text{\AA}$). \cite{Erickson1953, Birgeneau1971} The dominant magnetic interaction is the antiferromagnetic superexchange coupling between the $S=5/2$ spins of the Mn$^{2+}$ ions, between the eight next-nearest neighbors in MnF$_2$, and between the four nearest neighbors in the $ab$-plane in Rb$_2$MnF$_4$. A small anisotropy arising from dipolar interactions causes uniaxial spin alignment along the $c$-axis in both compounds below the respective N\'eel temperatures.\cite{Erickson1953, Birgeneau1971} Large single crystals of MnF$_2$ (Rb$_2$MnF$_4$) with a volume of $10\,\text{cm}^3$ ($3\,\text{cm}^3$) and mosaic spread of $0.44'$ ($0.99'$) FWHM were available from a previous experiment. \cite{bayrakci2013} The mosaic spreads were measured by $\gamma$-diffractometry at the $(200)$ reflections at room temperature. Elastic magnetic neutron scattering measurements of the antiferromagnetic order parameters of these samples (FIG.~\ref{fig:orderparameter}) yielded $T_N = 67.3\,\text{K}$ (MnF$_2$) and $38.4\,\text{K}$ (Rb$_2$MnF$_4$), in agreement with prior work. \cite{dietrich1969,schulhof1971,lee1998,leheny1999,christianson2001}

\begin{figure}[t]
	\includegraphics[width=\columnwidth]{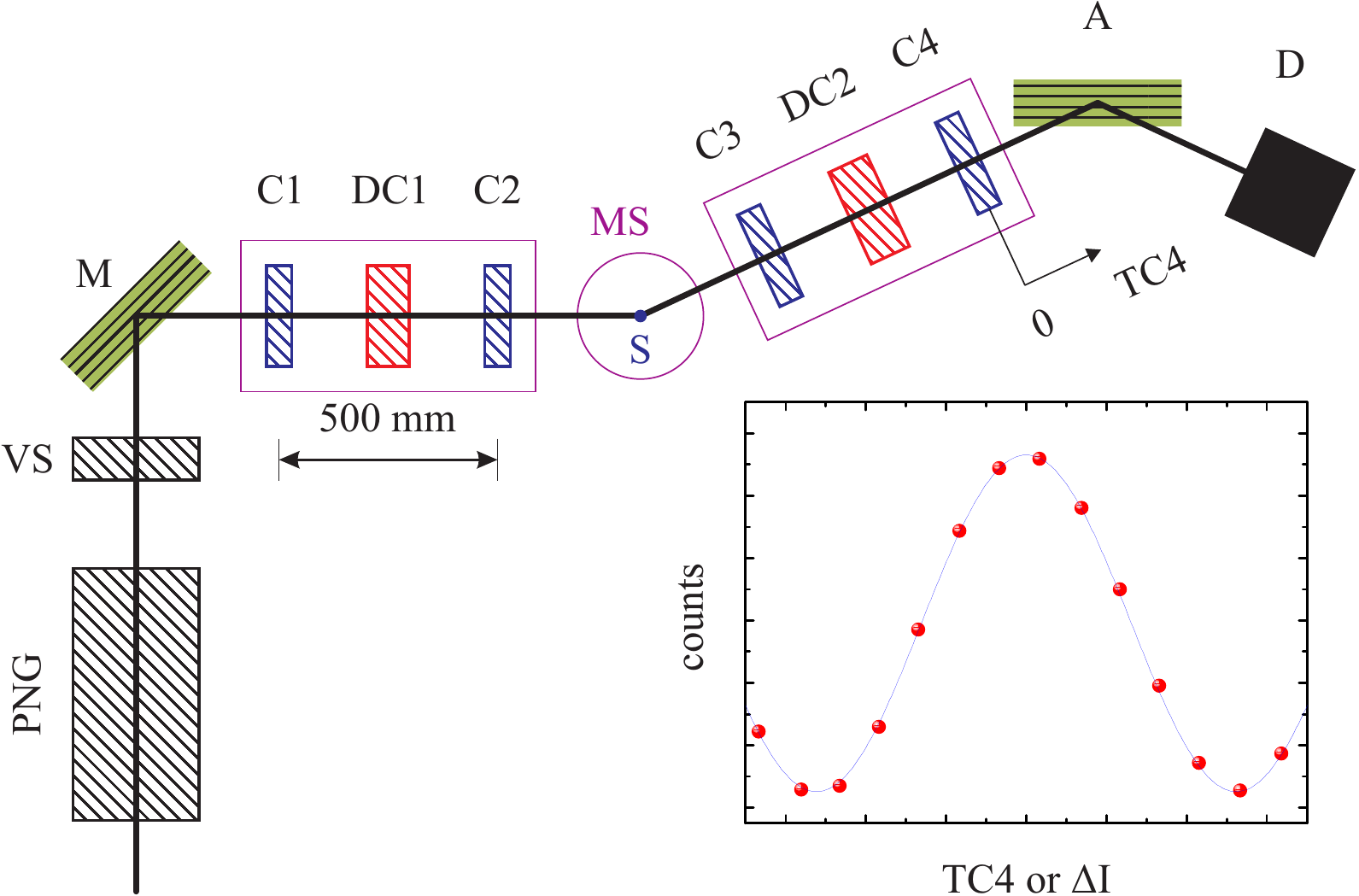}
	\caption{Schematic view of the TRISP spectrometer. M and A are the monochromator and analyzer, S is the sample and D the detector; VS is a velocity selector acting as higher order filter. The precession devices (PDs) are defined by pairs of radio-frequency coils (C1-C2) and (C3-C4) or by DC coils (DC1, DC2). To avoid spurious spin precession, mu-metal boxes and tubes (MS) enclosing the coils and the sample reduce external magnetic fields along the beam path to $<5\,\text{mG}$. Inset: Spin-echo raw data,  detector counts \emph{vs.} position TC4 of the coil C4 or \emph{vs.} difference in DC currents $\Delta I = I_2 - I_1$.}
	\label{fig:sketchTRISP}
\end{figure}

The experiments were conducted at the NRSE-TAS spectrometer TRISP \cite{keller2002} at the Maier-Leibnitz-Zentrum (MLZ) in Garching, Germany (FIG.~\ref{fig:sketchTRISP}). The crystals were mounted in a closed cycle cryostat in exchange gas in the $(H0L)$ (MnF$_2$) and $(HK0)$ (Rb$_2$MnF$_4$) scattering planes. The temperature was stable within $1\,\text{mK}$. Data were collected during several beam times with slightly varying crystal mounts. Consistent thermometry between these runs was ensured by measuring the intensities of magnetic Bragg reflections at the beginning of each run. $T_N$ is given by the maximum slope of the magnetic intensity, which varies by $\pm 0.07\,\text{K}$ between the individual experiments. We defined $T_N$ as the mean of all runs and adjusted the temperature scales by adding an offset such that the positions of maximum slope coincide. TRISP was operated with a graphite $(002)$ monochromator and a Heusler $(111)$ analyzer, with open collimation and scattering senses $SM=-1$, $SS=-1$, $SA=1$ at the monochromator, sample, and analyzer, respectively ($-1$ is clockwise). The data were collected at reciprocal lattice points corresponding to pure antiferromagnetic Bragg reflections. For the experiment on MnF$_2$ at $\bm{Q}=(300)$, we used an incident wave number $k_i=2.35\,\text{\AA}^{-1}$ with a TAS energy resolution $V=0.8\,\text{meV}$ (vanadium width, full width at half maximum, FWHM). For Rb$_2$MnF$_4$, $k_i$ was set to $2.5\,\text{\AA}^{-1}$ with $V=1.1\,\text{meV}$ at $\bm{Q}=(0.5\, 0.5\, 0)$.

In the following we highlight some features of the spin-echo technique that are relevant for the subsequent data analysis, and then discuss how spin-flip scattering affects the spin-echo signal and how we can discriminate between longitudinal and transverse fluctuations. The key components of a spin-echo spectrometer are two precession devices (PDs) bracketing the sample, either formed by uniform DC fields $\bm{B}_0$ (NSE) \cite{Mezei1972} or by pairs of radio-frequency (RF) spin-flip coils (NRSE), \cite{golub1987} where each RF coil incorporates both a RF field $B_1 \propto \cos(\omega_L t)$ and a DC field $\bm{B}_0$. Inside these two PDs the neutron spins undergo Larmor precession with frequency $\omega_L=\gamma B_0$, with $\gamma=2.916\,\text{kHz/Oe}$. The phase $\phi_\text{SE}$ \emph{vs.} frequency or current generated in the PDs was measured before the experiments and enters the fitting procedure as a known and fixed parameter. In the case of non-spin-flip scattering, the fields $\bm{B}_0$ of the two PDs are chosen to be opposite in sign, and the net precession angle $\Delta\phi_\text{SE}$ at the exit of the second region is a measure of the energy transfer $\hbar\omega$, with $\Delta\phi_\text{SE}=\omega\times\tau$. $\tau=m^2 \omega_L L /(\hbar^2 k_i^3)$ is the spin-echo time, $m$ is the neutron mass.

The polarization of the scattered beam is defined as $P=\langle\cos(\Delta\phi_\text{SE})\rangle$. In the case of non-spin-flip scattering,
\begin{equation}\label{eq:NsePol}
P(\tau) =P_0(\tau)\int S(\bm{Q}, \omega) R(\bm{Q}, \omega) \cos(\omega \tau) d\omega	
\end{equation}
where $S$ is the dynamic structure factor, and $R$ is the Gaussian TAS resolution function. $P_0$ is the spin-echo resolution function, including instrumental effects resulting from the beam divergence and from small field inhomogeneities of the PDs. \cite{Habicht2003} $S(\bm{Q},\omega)$ is usually Lorentzian in $\omega$ with a half-width-at-half-maximum $\Gamma$. In high-resolution spin-echo experiments, $\Gamma$ is usually much smaller than the energy width $V$ of $R(\bm{Q},\omega)$. In this case, the polarization is a simple exponential $P=\exp(-\Gamma \tau)$. At TRISP (FIG.~\ref{fig:sketchTRISP}), the RF coils can only be operated in a range $\tau_\text{min}\leq\tau\leq 20\times\tau_\text{min}$ with $\tau_\text{min}=4.09\,\text{ps}$ at $k_i=2.51\,\text{\AA}^{-1}$. For the present study, it was necessary to extend the $\tau$ range to zero by using small DC coils as PDs, which cover $0\leq\tau\leq 1.8\times\tau_\text{min}$ such that a good overlap with the RF-coils is given.

The polarization $P(\tau)$ is determined by detuning the precession phase of the second PD by a small amount $\Delta\phi_\text{off}$. The count rate $I(\Delta\phi_\text{off})$ varies sinusoidally with an amplitude $\propto P$:
\begin{equation}
I(\tau,\Delta\phi_\text{off}) = I_0[1+P(\tau)\cos(\Delta\phi_\text{off})]
\end{equation}
where $I_0$ is the mean intensity corresponding to $P=0$. In the operation mode using the RF coils, coil C4 is scanned along the beam direction, such that the widths of the two PDs differ by TC4 (FIG.~\ref{fig:sketchTRISP}). The phase offset is $\Delta\phi_\text{off}(\text{TC4}) = 2\pi \text{TC4} / \Delta \text{TC4}$, with the period $\Delta \text{TC4} = 2\pi\times\hbar k_f/(m\omega_L)$. In the low-resolution mode using the DC coils, the phase offset is $\Delta\phi_\text{off}(I_2)=2\pi(I_2-I_1)/\Delta I_2$, and the period is $\Delta I_2 = k_f/C_\text{coil}$ with $C_\text{coil}=49.9\,\text{\AA}^{-1}\text{A}^{-1}$ for the coils used at TRISP.

\begin{figure} [t]
	\includegraphics[width=0.7\columnwidth]{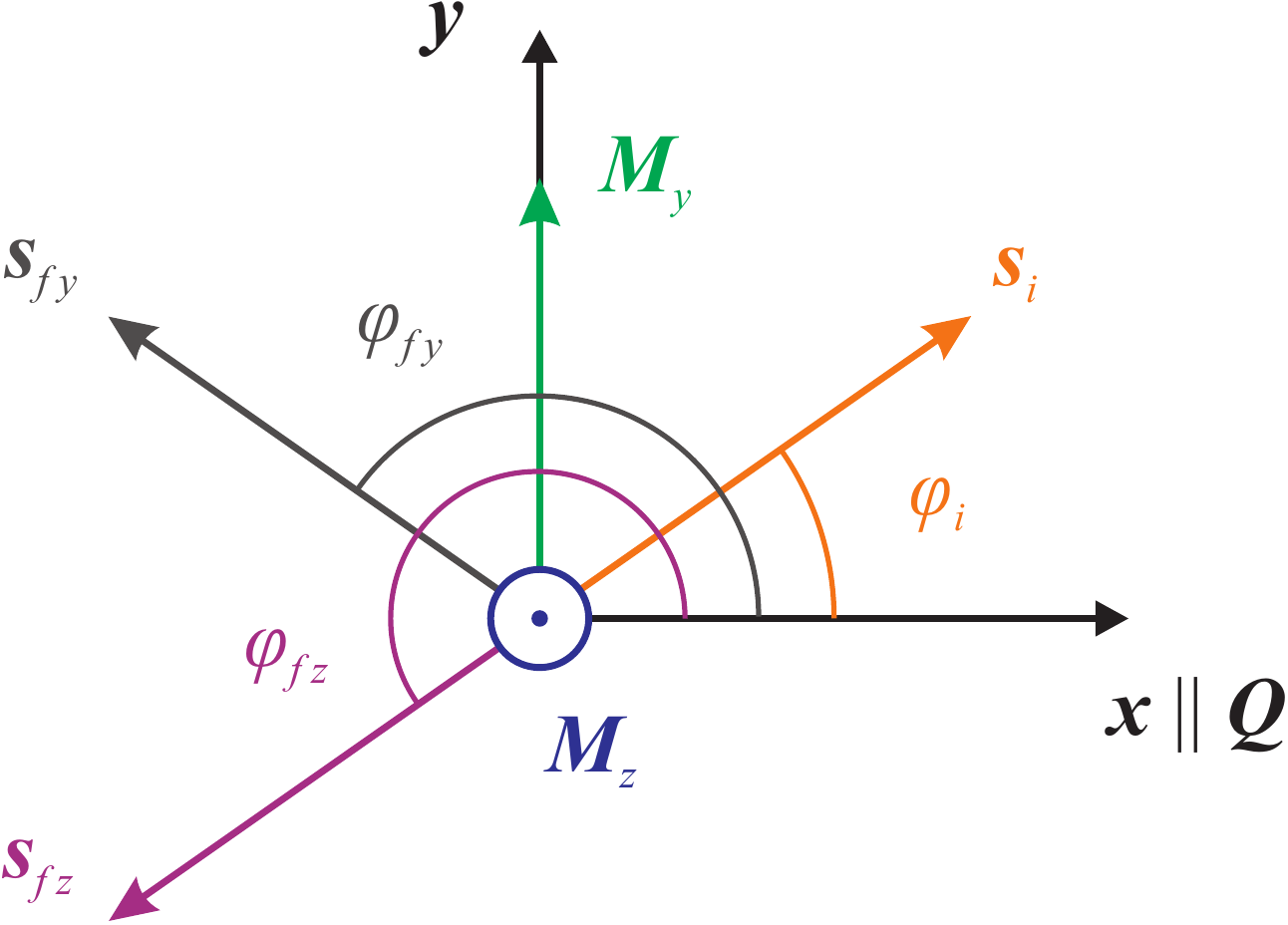}
	\caption{Spin flip processes at the sample. The spins $\bm{s}_i$ of the incident beam are spread within the horizontal $xy$-plane, where $x\parallel\bm{Q}$, $z$ is vertical. Only magnetic fluctuations $\bm{M}_y, \bm{M}_z \perp \bm{Q}$ contribute to the scattering cross section. The spin of the incident neutron $\bm{s}_i$ with Larmor phase $\phi_i$  is flipped to $\bm{s}_{fy}$ or $\bm{s}_{fz}$ by $\bm{M}_y$ or $\bm{M}_z$, respectively. The corresponding phases are $\phi_{fy} = \pi - \phi_i$ and $\phi_{fz} = \pi + \phi_i$.}
	\label{fig:SpinPhase}
\end{figure}

We now discuss the distribution of neutron spins at the sample, the spin-flip processes, and the influence of spin-flip scattering on the spin-echo signal. FIG.~\ref{fig:SpinPhase} shows the geometry of the neutron spin-flip processes arising from the magnetic fluctuations in the sample. At TRISP, the magnetic fields $\bm{B}_0$ of the precession devices are vertical ($z$-direction), and the neutron spins $\bm{s}$ precess in the horizontal $xy$-plane, also referred to as \emph{precession plane}. In the first PD, the neutron spins of an initially polarized beam accumulate a Larmor phase $\phi_i = m\omega_L L/(\hbar k_i)$, $m$ is the neutron mass. The phase varies due to the variation $\Delta k_i$ of the incident neutron wavenumber, so that $\Delta\phi_i = \Delta k_i/k_i\times\phi_i$. The width of the Gaussian $k_i$ distribution is $\Delta k_i/k_i =0.02$ for $k_i=2.51\,\text{\AA}^{-1}$. The incident beam is fully depolarized at the sample for $\Delta\phi_i>2\pi$, which happens for $\Delta\phi > 3\times10^2$ or $\tau>10\,\text{ps}$. This is in contrast to the usual 1D polarization analysis technique, \cite{moon1969} where at the sample all neutron spins are aligned in the same direction, parallel (or anti-parallel) to a guide field.

\begin{figure} [t]
	\includegraphics[width=\columnwidth]{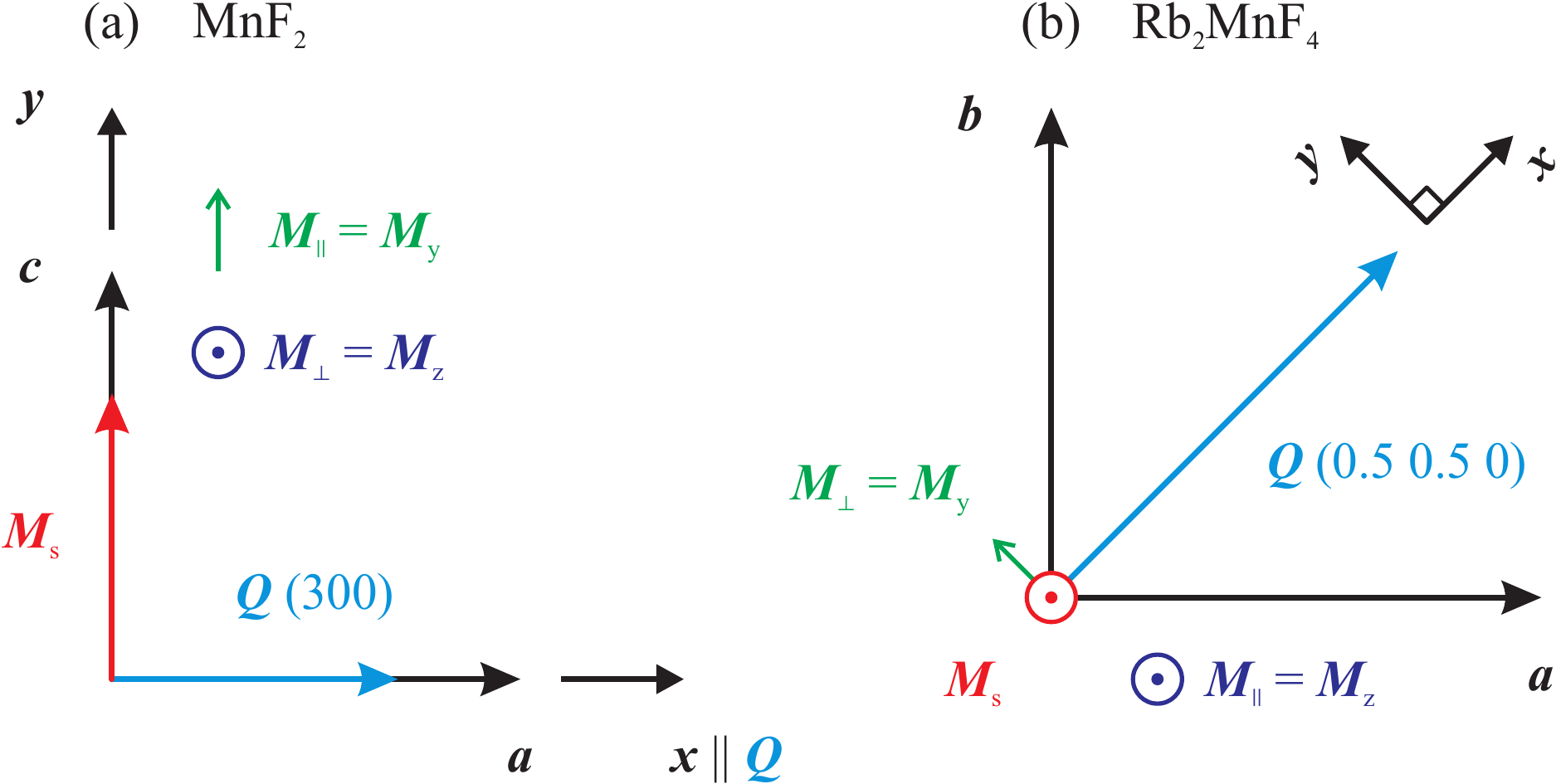}
	\caption{Spin fluctuations parallel and perpendicular to the sublattice magnetization $\bm{M}_s$ are referred to as longitudinal (labeled $\parallel$) and transverse (labeled $\perp$). In both MnF$_2$ and Rb$_2$MnF$_4$, $\bm{M}_s$ is parallel to the tetragonal $c$-axis. (a) In MnF$_2$, the $ac$-plane was aligned in the scattering plane, thus the longitudinal fluctuations $\bm{M}_\parallel$ are along $y$, and the transverse fluctuations $\bm{M}_\perp$ along $z$. (b) Rb$_2$MnF$_4$ was aligned in the $ab$-plane with $\bm{M}_\parallel$ along $z$ and $\bm{M}_\perp$ along $y$.}
	\label{Fig:ScattGeom}
\end{figure}

The relation between the coordinates $xyz$ and the longitudinal and transverse spin fluctuations $\bm{M}_\parallel$ and $\bm{M}_\perp$ is shown in FIG.~\ref{Fig:ScattGeom}. $\bm{s}_i$ undergoes a $\pi$ flip around the respective component of $\bm{M}$, such that $\bm{M}_z$ flips $\bm{s}_i$ to $\bm{s}_{fz}$ with $\phi_{fz}=\phi_i+\pi$. \cite{moon1969} $\bm{M}_y$ flips $\bm{s}_i$ to $\bm{s}_{fy}$ with $\phi_{fy}=-\phi_i + \pi$ and thus inverts the sign of $\phi_i$. This is an effective sign inversion of the field $\bm{B}_0$ in the first PD. The echo condition is fulfilled, that is, the Larmor phase of the first PD is inverted in the second one, if the fields $\bm{B}_0$ of the two PDs are anti-parallel for $\bm{M}_z$ and parallel for $\bm{M}_y$. The neutron spins scattered by the component of $\bm{M}$ not fulfilling the echo conditions effectively precess with the same sign in both PDs. They are depolarized if their phase is spread by more than $2\pi$ at the exit of the second region. Following the previous argument about depolarization of the neutron beam at the sample, this happens for $\tau>5\,\text{ps}$.

\section{Data analysis}
	
\begin{figure} [t]
   \includegraphics[width=\columnwidth]{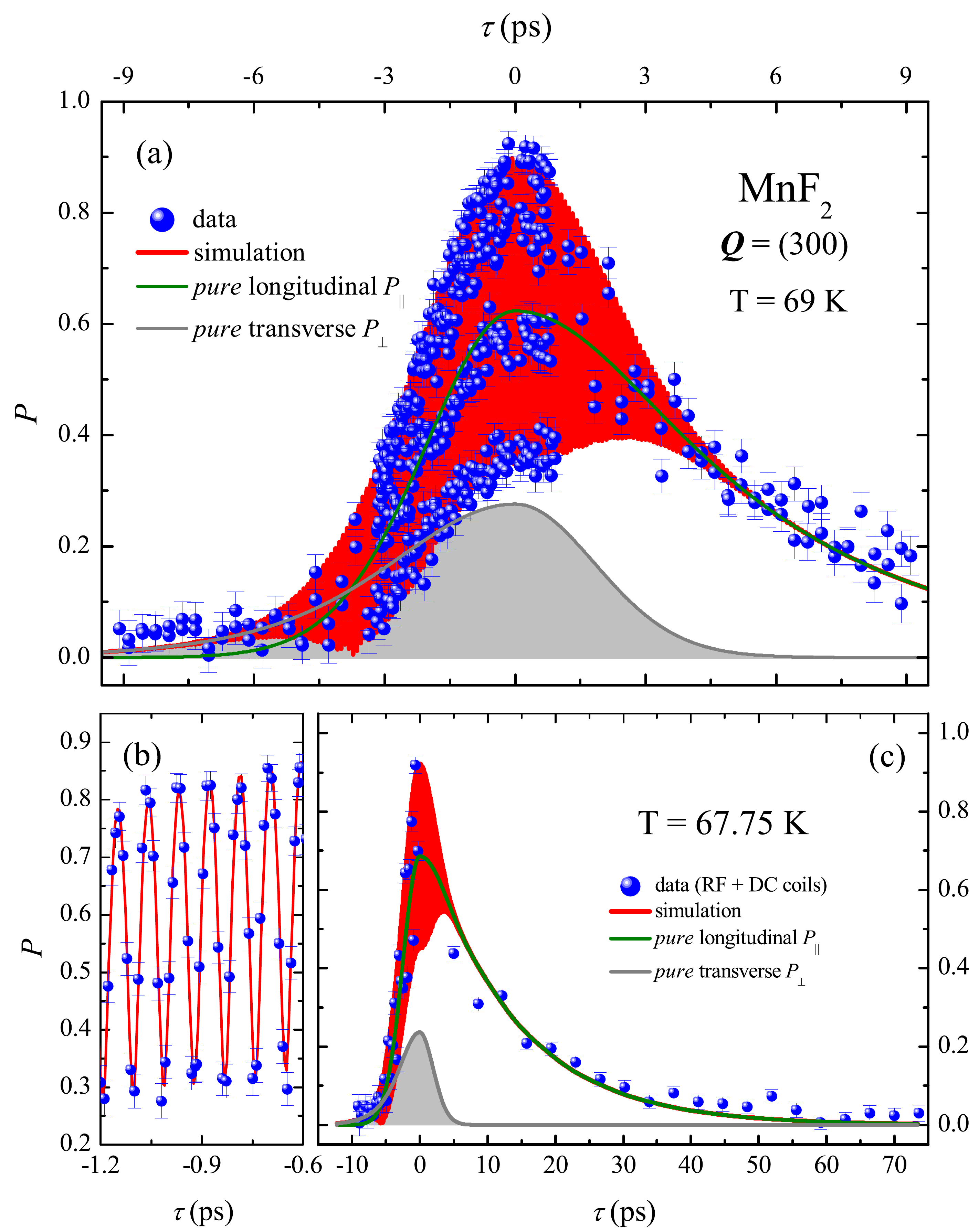}
	\caption{Sample echo data of critical scattering in MnF$_2$ and fit with the model described in the text at $\bm{Q}=(300)$ at (a) $T=69\,\text{K}$ and (c) $T=67.75\,\text{K}$, where $T_N=67.3\,\text{K}$. (a) and the zoom (b) show the fast oscillation of the polarization resulting from the interference of scattering by $\bm{M}_y$ and $\bm{M}_z$. A positive (negative) sign of $\tau$ corresponds to parallel (anti-parallel) $\bm{B}_0$. The lines $P_\parallel$ and $P_\perp$ show the contribution of the longitudinal and transverse fluctuations to the polarization, where the peaks of these curves are proportional to the integrated intensities.}
	\label{Fig:MnF2.sample}
\end{figure}

FIG.~\ref{Fig:MnF2.sample} shows typical data $P(\tau)$ for MnF$_2$ at the pure antiferromagnetic Bragg point $\bm{Q}=(300)$. A prominent feature of the data is the fast oscillation of the polarization, which is displayed as red area in panel (a) and resolved in the zoomed version in panel (b). These oscillations result from the $\tau$-dependent phase $\phi_{fz} - \phi_{fy} = 2\phi_i(\tau)$ between $\bm{s}_{fz}$ and $\bm{s}_{fy}$ (FIG.~\ref{fig:SpinPhase}), with $\phi_i[\text{rad}]= 3.15\times\tau[\text{ps}]\times (k_i[\text{\AA}^{-1}])^2$. For positive $\tau$ (parallel $\bm{B}_0$ configuration), only the spins $\bm{s}_{fy}$ obey the echo condition, whereas the spins $\bm{s}_{fz}$ are depolarized with increasing $\tau$, such that the oscillation amplitude decreases. For negative $\tau$ (anti-parallel $\bm{B}_0$) $\bm{s}_{fz}$ fulfill the echo condition and the remaining polarization of $\bm{s}_{fy}$ generates the oscillations. At large $\tau$ beyond the oscillation regime, $P(\tau$) can be modeled by Eq.~(\ref{eq:NsePol}). Thus the asymmetry in the decay between $\tau>0$ and $\tau<0$ indicates $\Gamma_\parallel \ll \Gamma_\perp$, both for $T=69\,\text{K}$ and $T=67.75\,\text{K}$. The smaller oscillation amplitude at $T=67.75\,\text{K}$ arises from a larger relative intensity of the neutrons scattered by the longitudinal fluctuations $\bm{M}_\parallel$.

To devise an analytic model describing the entire data set independent of approximations, we implemented a numerical calculation of $P(\tau)$ based on a ray-tracing model, which traces the spin of individual neutrons in the PDs and in the scattering process. We first assume $S(\bm{Q}, \omega)$ to be independent of $\bm{Q}$ within the small momentum range defined by the TAS resolution ellipsoid. The small effect of the finite momentum resolution is discussed below. The energy dependence is modeled as Lorentzian:
\begin{equation} \label{eq:S.omega}
	S(\omega) = \frac{A \Gamma_\parallel}{\Gamma_\parallel^2 + \omega^2} +
	\frac{(1-A) \Gamma_\perp}{\Gamma_\perp^2 + \omega^2}
\end{equation}
where $A$ and $(1-A)$ are proportional to the integrated intensities scattered by the longitudinal and transverse fluctuations, respectively. We also tested a second model of $S(\omega)$ allowing for two split Lorentzian modes $\Gamma_\perp^\pm$ for $T<T_N$ similar to the observations in the 3DHA RbMnF$_3$, \cite{coldea1998} but we obtained no improvement of the fit quality. All our data are consistent with the model in Eq.~(\ref{eq:S.omega}). Further we assume a Gaussian distribution of $k_i$. The resolution function of the TAS, $R(\omega)$, is modeled as a Gaussian, and the width is taken as the vanadium width $V$ determined experimentally. The parameters assigned to each neutron are $k_i$, a 3D polarization vector, and a probability $p$ to find a neutron in this state. In contrast to usual ray tracing packages, the choices of $k_i$, the scattering process ($\parallel$ or $\perp$ fluctuations), and the energy transfer $\omega$ are not based on random numbers, but on equally spaced grids.  This avoids the statistical noise introduced by random numbers, which disturbs the minimization algorithm of the fitting routine. \cite{James1994} The energy transfers $\omega_{\parallel, \perp}$ are taken in a band $\pm10 \Gamma_{\parallel,\perp}$ with about $200$ points to avoid cutting of the Lorentzian wings. The polarization $P(\tau, \Delta\phi_\text{off}, k_i, A, \Gamma_\parallel, \Gamma_\perp)$ calculated within this model is in excellent agreement with our entire data set (FIG.~\ref{Fig:MnF2.sample}).

We now discuss the effect of the finite momentum resolution defined by the TAS resolution ellipsoid $R(\bm{Q},\omega)$. The data of the present experiments were taken at magnetic Bragg reflections $\bm{G}$, where $\bm{q} = \bm{G} - \bm{Q}$ and $S(\bm{q}, \omega)$ vary within the region defined by $R$. To estimate the effect on the linewidth measured by spin-echo, we calculated the polarization from Eq.~(\ref{eq:NsePol}), where the $R(\bm{Q},\omega)$ was calculated with matrix elements corresponding to the spectrometer configurations. \cite{cooper1967} $S(q,\omega$) was taken from Refs.~\onlinecite{schulhof1971,christianson2001}. The resulting additional broadening of the linewidth is roughly independent of temperature for $T \geq T_N$ and amounts to about $5\,\mu\text{eV}$ in MnF$_2$ and $0.8\,\mu\text{eV}$ in Rb$_2$MnF$_4$. The reason for the larger value in MnF$_2$ is the relaxed vertical resolution $Q_z$, which has no effect in the 2D spin system of Rb$_2$MnF$_4$.

\section{Results and Discussion}
\subsection{MnF$_2$}

\begin{figure}
	\includegraphics[width=\columnwidth]{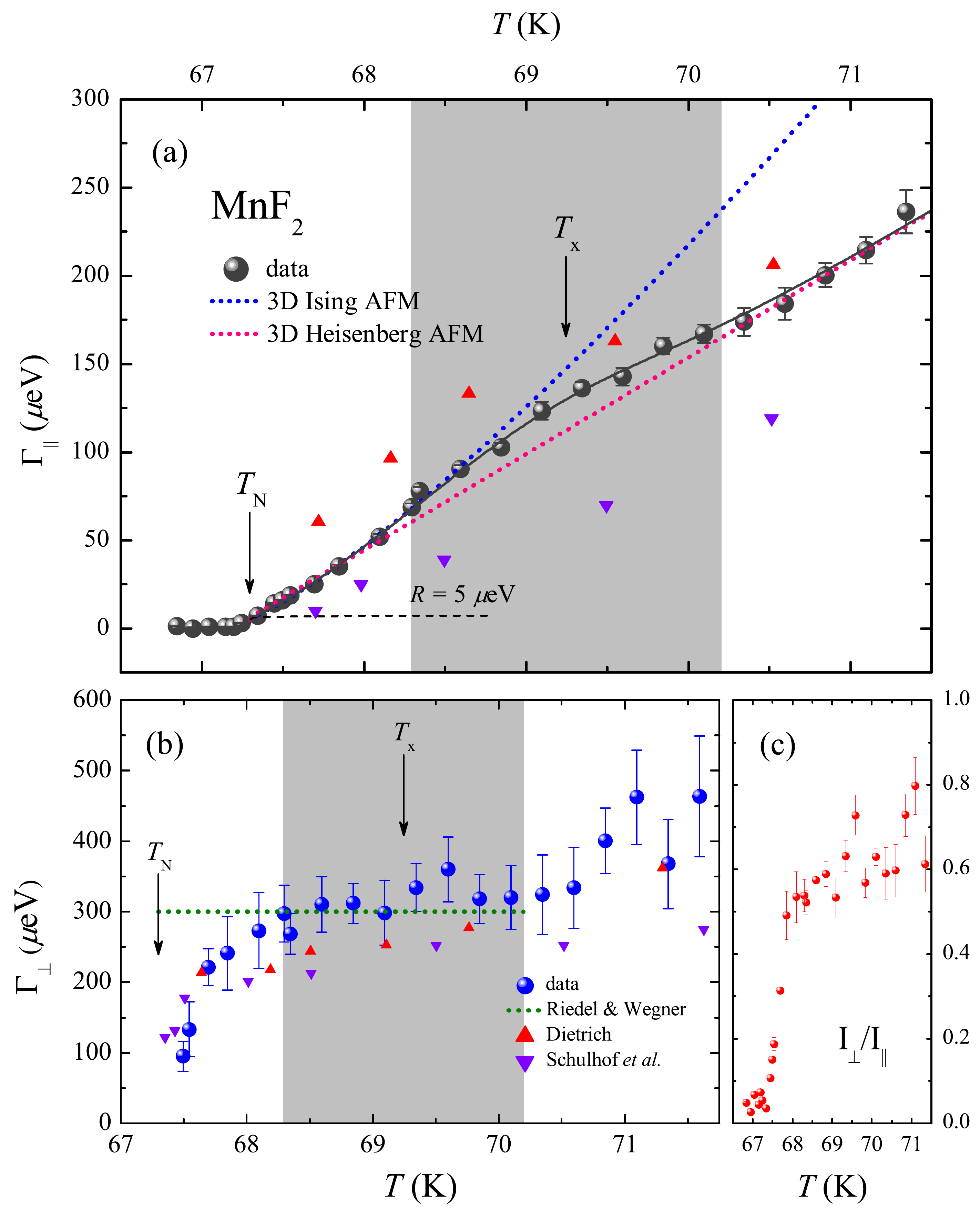}
	\caption{ Temperature dependence of $\Gamma_{\parallel, \perp}$ in MnF$_2$ at $\bm{Q}=(300)$. (a) $\Gamma_\parallel$ shows a crossover from 3D Ising to 3D Heisenberg critical scaling, where the gray band indicates the crossover region centered at $T_x$. $R=5\,\mu\text{eV}$ is the broadening due to the finite momentum resolution. (b) $\Gamma_\perp$ and data from early TAS experiments. \cite{dietrich1969,schulhof1971} The green dotted line shows the calculated $\Gamma_\perp$. \cite{riedel1970} (c) Ratio of integrated intensities $I_\perp / I_\parallel$. Close to $T_N$, $I_\parallel$ is much stronger. For $T > T_x$ in the 3DHA region, $I_\perp / I_\parallel$ is growing within the experimental temperature range and approaches unity for $T\gg T_N$.}
\label{Fig:MnF2_Fig}
\end{figure}

FIG.~\ref{Fig:MnF2_Fig} shows the longitudinal ($\Gamma_\parallel$) and transverse ($\Gamma_\perp$) linewidths at $\bm{Q}=(300)$ extracted with the aforementioned model. Only the longitudinal fluctuations show critical behavior around $T_N$ while the transverse ones evolve continuously, as expected based on the spin anisotropy and the uniaxial order parameter. \cite{schulhof1971} $T_N=67.29\,\text{K}$ was determined from the maximum slope of the intensity of the magnetic $(300)$ Bragg reflection (FIG.~\ref{fig:orderparameter}), and is in agreement with the literature values. \cite{Bruce1981} The measured linewidth $\Gamma_\parallel(T=T_N)=5\,\mu\text{eV}$ is larger than the intrinsic spectrometer resolution ($<1\mu\text{eV}$) and agrees with the value calculated above by taking the finite $\bm{Q}$ resolution into account. According to the dynamical scaling prediction, \cite{halperin1969} $\Gamma_\parallel$ follows a power law
\begin{equation}\label{eq:GammaParallel}
\Gamma_\parallel(T)=A_\parallel t^{z\nu} \propto\kappa_\parallel^z
\end{equation}
where $A_\parallel$ is a normalized amplitude, $t=T/T_N-1$ is the reduced temperature, and $\kappa=\xi^{-1} \propto t^\nu$ is the inverse correlation length.

The $\Gamma_\parallel (T)$ data in FIG.~\ref{Fig:MnF2_Fig}(a) clearly deviate from a single power law in the shaded region around $T=69\,\text{K}$. We thus performed separate fits to the regions below and above 69 K. The blue dotted line fits the data in the range $T_N<T<1.01T_N$, with $z\nu = 1.25(2)$. With the exponent $\nu_\text{3DIA}=0.6301$ predicted for 3DIA scaling, \cite{pelissetto2002} we obtain $z=1.98(3)$, which matches the $z_\text{3DIA}=2$ expected for this universality class within the experimental error. \cite{hohenberg1977} 3DHA scaling in this temperature range can be excluded: dividing $z\nu$ by $\nu_\text{3DHA}=0.7112$ predicted for the 3DHA \cite{campostrini2002} results in $z=1.77$, inconsistent with $z_\text{3DHA}=1.5$ predicted for the 3DHA. \cite{hohenberg1977} For $T>1.04T_N$, the red dotted curve corresponds to an exponent $z\nu=1.02(3)$. Dividing by $\nu_\text{3DHA}$ gives $z=1.43(5)$, close to 3DHA scaling, whereas the $z=1.62(4)$ obtained with $\nu_\text{3DIA}$ is inconsistent with the theoretical $z_\text{3DIA}=2$. Thus the data $\Gamma_\parallel(T)$ show a crossover from 3DIA close to $T_N$ to 3DHA scaling for $T\gg T_N$. This relative amplitudes $A_{\parallel \text{3DIA}}/A_{\parallel \text{3DHA}}=3.0$ resulting from the fits are in good agreement with the value $3.1$ predicted by \citealt{riedel1970} who extended the dynamical scaling theory to anisotropic systems.

For a quantitative description of the crossover region of $\Gamma_\parallel(T)$ we use the phenomenological expression
\begin{equation}\label{eq:crossover}
\Gamma(T)=[1-H(T-T_x)]\Gamma_\text{Ising} + H(T-T_x)\Gamma_\text{Heis}
\end{equation}
where $H(T-T_x) = 1/2+1/2\tanh[\gamma(T-T_x)]$ is a slowly varying function symmetrically centered at a crossover temperature $T_x$. The transition width $\gamma$ is defined as the region $0.1<H<0.9$ describing the crossover temperature range. A fit of Eq.~(\ref{eq:crossover}) to our data gives a crossover region $1.01T_N<T<1.04T_N$ with $T_x=69.2(1)\,\text{K}$ or $t_x=0.029(1)$. \citealt{pfeuty1974} predicted such a crossover for antiferromagnetic 3DIA to 3DHA scaling for $t_x=\alpha_I^{0.8}$, where $\alpha_I=H_A/H_E$ is the ratio of anisotropy and exchange fields in the spin Hamiltonian. $\alpha_I=0.016$ for MnF$_2$ gives $t_x=0.036$, in good agreement with our experimental result. \cite{johnson1959}

\citealt{schulhof1971} pointed out that their result for MnF$_2$ favors the value $z=1.5$, consistent with 3DHA scaling, whereas the static exponents $\nu$ and $\gamma$ agree with the 3DIA model. They argued that the reason for this discrepancy might be the small range in momentum $\bm{q}$ where the crossover is visible in $\Gamma_\parallel$. \citealt{riedel_prl1970,riedel1970} introduced the parameter $\kappa_\Delta=\kappa_\parallel(t_x,q=0)$ defining the crossover between isotropic and anisotropic regions in momentum space, with $\kappa_\parallel^{2}+q^{2}=\kappa_{\Delta}^{2}$. They estimate $\kappa_\Delta=0.054\text{\AA}^{-1}$ for MnF$_2$, corresponding to $T_x\approx T_N+2\,\text{K}$, close to the observation in the present work. \citealt{Frey1994} obtained a similar value of $\kappa_\Delta=0.06\,\text{\AA}^{-1}$ in a calculation of the critical dynamics taking dipolar interactions into account. Since the linewidths $\Gamma_\parallel$ in this region were too narrow to be resolved by TAS, the crossover of the dynamical exponent $z$ was missed. For the strongly anisotropic antiferromagnet FeF$_2$, both $t_x=0.45$ and $\kappa_\Delta=0.29\,\text{\AA}^{-1}$ are larger, such that the TAS experiment covered the 3D Ising region close to $T_N$ without observing the crossover to Heisenberg dynamic scaling. \cite{hutchings1972}

The width $\Gamma_\perp$ of the transverse fluctuations is shown in FIG.~\ref{Fig:MnF2_Fig}(b) in comparison with TAS data from Refs.~\onlinecite{dietrich1969,schulhof1971}. We observe a rapid increase of $\Gamma_\perp$ between $T_N$ and the lower bound of the crossover region at $1.01T_N$, where $\Gamma_\perp$ saturates at $\sim 0.3\,\text{meV}$. Calculations predicted this saturation value, corresponding to $z_\perp=0$. \cite{riedel1970,Frey1994,tsai2003} But $\Gamma_\perp$ is expected to stay constant in the broad range $T_N<T<T_x$, which contradicts both our data and the results of the early TAS experiments. $\Gamma_\perp$ increases beyond the crossover region ($T>1.04T_N$), as expected for the 3DHA. The error bars increase at high temperature, because the wings of the Lorentzian line are cut by the transmission function $R(\omega)$ of the NRSE-TAS spectrometer ($\sim 0.8\,\text{meV}$ FWHM). Thus the data quality does not allow fitting of a critical exponent and quantitative confirmation of 3DHA scaling of $\Gamma_\perp$ for $T \gg T_N$.

\subsection{Rb$_2$MnF$_4$}

\begin{figure*}[t]
	\includegraphics[width=2\columnwidth]{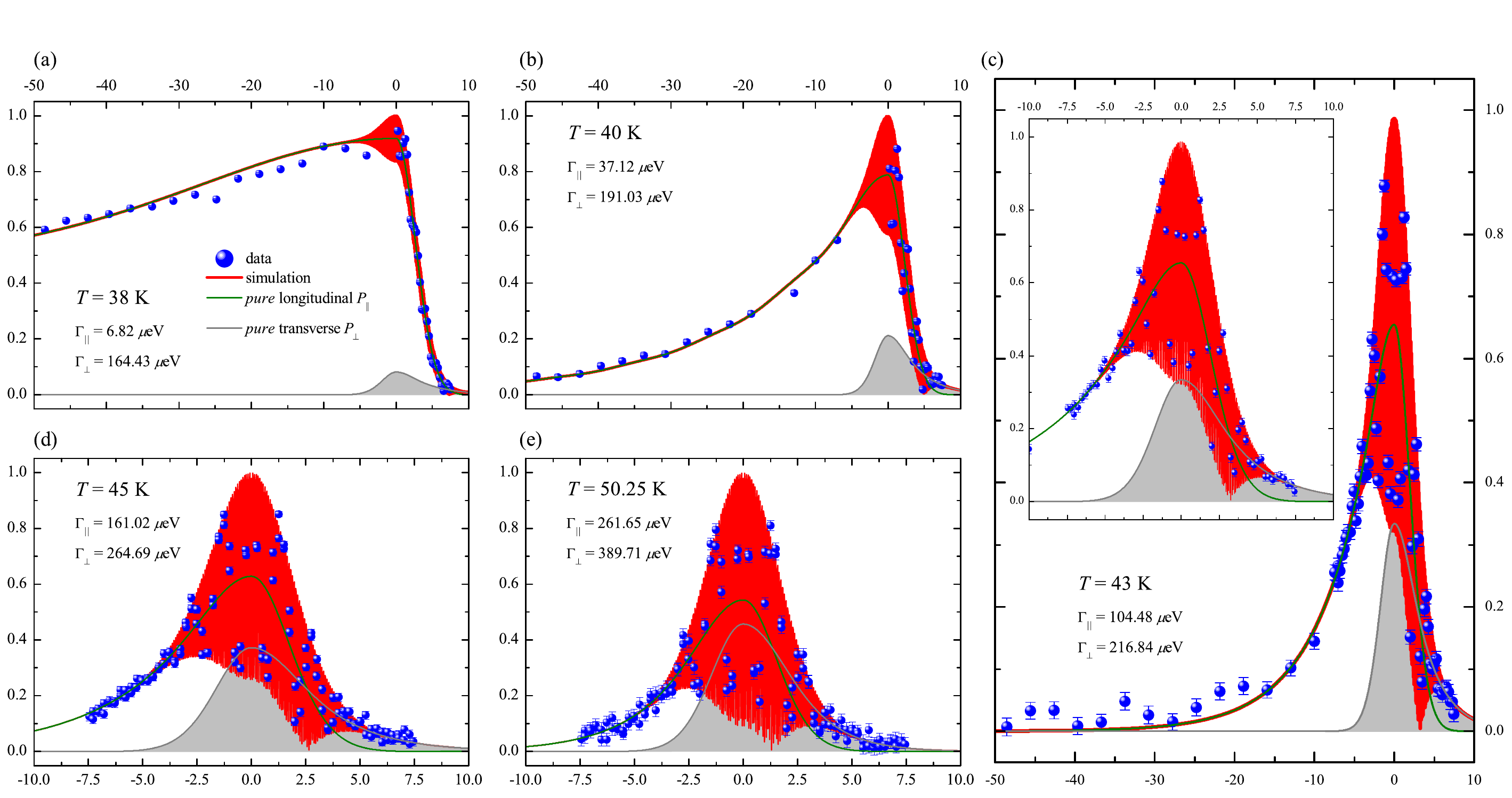}
	\caption{Spin-echo polarization $P$ \emph{vs.} spin echo time $\tau[ps]$ at selected $T>T_N=37.6\,\text{K}$ measured at $\bm{Q}=(0.5\,0.5\,0)$ in Rb$_2$MnF$_4$. The fit (red line) using the model discussed in the text reproduces the fast oscillations in the region of small $\tau$, displayed as red areas. The components of the polarization $P_\parallel$ and $P_\perp$ corresponding to longitudinal and transverse fluctuations are show in green and gray, the peak values of these two curves are proportional to the relative intensities $I_{\parallel,\perp} = I_{\parallel,\perp} /(I_\parallel+I_\perp)$. $I_{\parallel,\perp}$ are the integrated intensities. The inset of (c) shows a zoom to the $T=43\,\text{K}$ data.}
	\label{Fig:RbMnF_ex}
\end{figure*}

Spin-echo data of critical fluctuations in Rb$_2$MnF$_4$ were measured at $\bm{Q}=(0.5\,0.5\,0)$, a pure magnetic Bragg reflection in the antiferromagnetically ordered state. The intensity of this reflection is shown in FIG.~\ref{fig:orderparameter}(b) as a function of temperature. The sharp peak results from the longitudinal critical scattering and defines $T_N=37.6\,\text{K}$, \cite{Ikeda1987} close to values from the literature. \cite{lee1998,leheny1999,christianson2001} Representative spin-echo data are shown in FIG.~\ref{Fig:RbMnF_ex}. Both fluctuation components $\bm{M}_\parallel$ and $\bm{M}_\perp$ contribute to the scattering cross section. According to FIG.~\ref{Fig:ScattGeom}(b), $\bm{M}_\parallel$ ($\bm{M}_\perp$) is perpendicular (parallel) to the scattering plane $xy$ and fulfills the spin-echo condition for negative (positive) $\tau$ corresponding to anti-parallel (parallel) $\bm{B}_0$. Close to $T_N$, the intensity of the longitudinal fluctuations dominates, and the transverse ones have nearly no effect on the signal. $\Gamma_\parallel$ is small, so that for $\tau<0$ (where the $\bm{M}_\parallel$ scattering fulfills the spin-echo condition) the polarization decays slowly. Upon heating (FIG.~\ref{Fig:RbMnF_ex}(b)-(e)) the intensity ratio $I_\perp / I_\parallel$ approaches unity, as expected for isotropic spin fluctuations, and $\Gamma_\parallel$ increases rapidly, leading to a faster decay of $P(\tau < 0)$. $\Gamma_\perp$ is rather large at $T_N$ and evolves more smoothly upon heating, so that $P(\tau>0)$ shows less variation with temperature.

\begin{figure}[t]
	\includegraphics[width=\columnwidth]{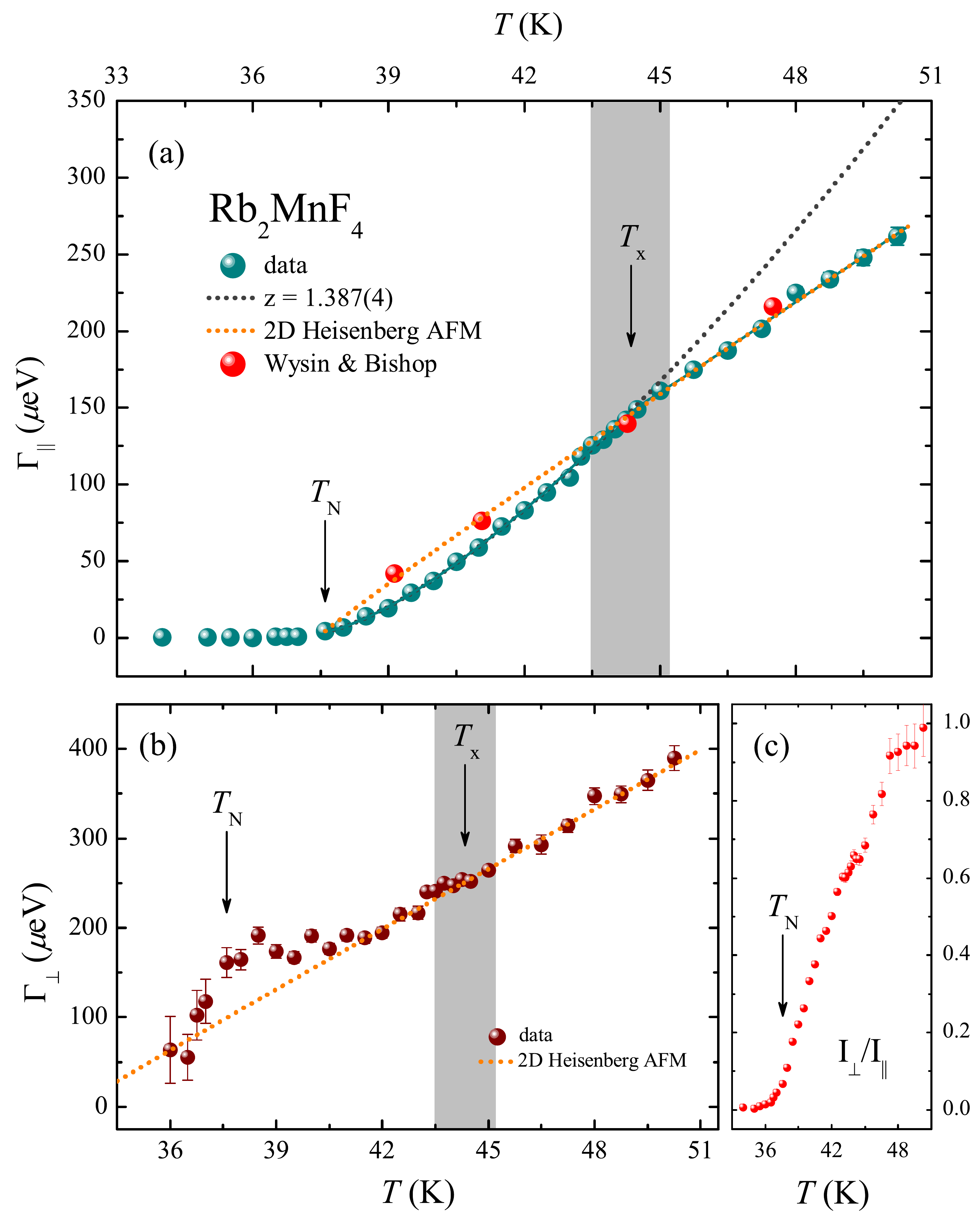}
	\caption{Linewidth \emph{vs.} $T$ of the critical fluctuations in Rb$_2$MnF$_4$ at $\bm{Q}=(0.5\,0.5\,0)$. (a) $\Gamma_\parallel$ shows a crossover in the scaling at $T_x=44.3\,\text{K}$, where above $T_x$ 2DHA is observed. (b) $\Gamma_\perp$ (c) Ratio of integrated intensities $I_\perp/I_\parallel$.}
	\label{Fig:Rb2MnF4.Gamma}
\end{figure}

FIG.~\ref{Fig:Rb2MnF4.Gamma}(a) shows the linewidth $\Gamma_\parallel$ of the longitudinal fluctuations. The broadening of $\Gamma_\parallel$ sets in about $0.6\,\text{K}$ below $T_N$ and reaches $4.3\,\mu\text{eV}$ at $T_N$. This value is larger than the calculated resolution of $\sim1.6\,\mu\text{eV}$, including $\sim0.8\,\mu\text{eV}$ intrinsic resolution and $0.8\,\mu\text{eV}$ broadening from the finite momentum resolution. The latter value was calculated assuming 2D correlations in the $xy$-plane, such that the vertical momentum resolution $Q_z$ has no effect. Very close to $T_N$, where the fluctuations leading to the 3D order also must reflect 3D correlations, such that the finite $Q_z$ resolution should become relevant. However, this temperature regime is very narrow, and the resolution correction should be insignificant in the range of reduced temperatures we are probing. \cite{birgeneau1970} Nonetheless, we note that the observed width at $T_N$ is very similar to the one in MnF$_2$ at $T_N$, where it most likely arises from the 3D spin correlations in conjunction with the poor vertical resolution. It is also similar to the residual linewidth of magnons at $T = 3$ K, deep in the N\'eel state of Rb$_2$MnF$_4$, which could be attributed to the effect of structural and/or magnetic domain boundaries. \cite{bayrakci2013} Further work is required to determine whether the small linewidth at $T_N$ arises from an unidentified resolution effect or from intrinsic properties of the sample such as residual disorder. In the following analysis, we subtract this contribution from the temperature dependent $\Gamma_\parallel$ data.

$\Gamma_\parallel$ shows a change in slope around $44\,\text{K}$. From the dipolar anisotropy one expects a crossover from 2DIA scaling for $T \sim T_N$ to 2DHA behavior for $T \gg T_N$. Such a crossover was observed \cite{lee1998} for the correlation length $\xi_\parallel$ close to $T_x = 1.2 T_N$. This value of $T_x$ was calculated for an anisotropy parameter $\alpha_I=0.0047$ extracted from the spin wave dynamics. \cite{deWijn1973,cowley1977,hamer1994} Fitting the power law $\Gamma_\parallel(t)$ of Eq.~(\ref{eq:GammaParallel}) in the range $T_N < T < 1.16 T_N$ gives an exponent $z\nu = 1.387(4)$. This value depends only weakly on the choice of the fitting range; removing two data points at the upper or lower boundary changes the result within the error bar. Using the exponent $\nu_\text{2DIA} = 1$ predicted for 2DIA scaling, \cite{onsager1944} we obtain $z=1.387(4)$, clearly different from the $z_\text{2DIA}=1.75$ predicted for the 2D Ising model. \cite{mazenko1981}  Other simple models, such as the 3DIA, also do not fit. With $\nu_\text{3DIA} = 0.6301$, we obtain $z = 2.201(6)$, different from the predicted $z_\text{3DIA}=2$. This means that our linewidth data close to $T_N$ are not consistent with the 2DIA behavior observed for the correlation length $\xi_\parallel$. \cite{lee1998} Such a deviation from 2DIA scaling was also observed for the static exponent $\beta$ for $T<T_N$. \cite{birgeneau1970}

A possible reason for the unexpected scaling of $\Gamma_\parallel$ is the the dipolar interaction, which is the major contributor to the magnon gap in the antiferromagnetically ordered state and can affect the universality class by virtue of its long spatial range. Based on theoretical considerations, Refs.~\onlinecite{Aharony1973,Frey1994} argued that the long-range nature of the dipolar forces should have no effect on the correlation length in antiferromagnets, but that the critical dynamics are modified by additional damping processes, especially in the limit of small $\bm{q}$ and close to $T_N$. In 3D antiferromagnets such as MnF$_2$, the critical regime in which the long-range character of the dipolar interaction significantly affects the critical scaling is expected to be small. \cite{Fischer1990} Indeed, our investigation of MnF$_2$ did not uncover any evidence of such an effect. For the 2D case, a stronger influence of the long range character is expected, \cite{Frey1994} but to the best of our knowledge a calculation of the critical dynamics of a 2D antiferromagnets with dipolar interactions has not yet been reported. It is interesting to note that the critical exponent in a magnetic field $H$ close to the bicritical point in the $H-T$ phase diagram of Rb$_2$MnF$_4$, $z = 1.35 \pm 0.02$, \cite{christianson2001} is identical to ours within the experimental error. This suggests that the magnetic field does not close the damping channels actuated by the dipolar interaction.

For $T\gg T_N$ the impact of the anisotropy decreases, and the fluctuations are expected to follow the 2DHA model which exhibits magnetic long range order only for $T \rightarrow 0$. \cite{mermin1966} The correlation length $\xi_\text{2DHA}$ for the pure $S=5/2$ 2DHA has been calculated by Cuccoli {\it et al.}, \cite{cuccoli1996,cuccoli1997} and the influence of the small spin-space anisotropy can be described by the mean-field expression $\xi_\text{eff}$: \cite{keimer1992}
\begin{equation}\label{eq:xi.effective}
	\xi_\text{eff}(\alpha_I, T) =\frac{\xi_\text{2DHA}}{\sqrt{1-\alpha_I\xi^2_\text{2DHA}(T)}}
\end{equation}
The effective correlation length $\xi_\text{eff}^{-1}$ is plotted in the inset of FIG.~\ref{fig:orderparameter}(b). Fitting the expression $\Gamma_\parallel(t) = A_\parallel \times \xi_\text{eff}^{-z_\parallel}(t)$ to the data $\Gamma_\parallel$ at $T>1.20T_N$ gives $z_\parallel = 0.96(4)$, in agreement with the prediction $z=1$ for the 2DHA. \cite{hohenberg1977} This result also agrees with a numerical simulation of $\Gamma_\parallel$ by \citealt{wysin1990}, also shown in FIG.~\ref{Fig:RbMnF_ex}(a), and with experimental results on a 2DHA model compound with $S=1/2$. \cite{kim2001} Finally we analyzed the entire data set $\Gamma_\parallel(T>T_N)$ with the crossover function in Eq.~(\ref{eq:crossover}). The resulting $T_x=44.3(4)$ ($t_x=0.179$) is slightly smaller than the predicted value, and the width of the crossover region is $1.7\,\text{K}$.

The linewidth $\Gamma_\perp$ of the transverse fluctuations is plotted in FIG.~\ref{Fig:Rb2MnF4.Gamma}(b). $\Gamma_\perp$ is nonzero at $T_N$,  forms a plateau with $z_\perp\sim0$ between $T_N$ and $T_x$, and grows continuously for $T>T_x$. In the 2DHA regime observed for $\Gamma_\parallel(T>T_x)$, it is expected that $\Gamma_\perp(t)=\Gamma_\parallel(t)$. \cite{riedel_prl1970}. It was pointed out that the effective N\'eel temperatures for the longitudinal and transverse fluctuations $T_\parallel$ and $T_\perp$ are different, \cite{moriya1962} such that $t=T/T_{\parallel,\perp}-1$. $T_N$ relevant for the magnetic ordering is the larger $T_\parallel$. We then fit $\Gamma_\perp = A_\perp \times \xi_\text{eff}^{-z_\perp}$ to the data $\Gamma_\perp(T>T_x)$ assuming $A_\perp=A_\parallel$, where the latter is known from the scaling of $\Gamma_\parallel$. This fit gives $T_\perp=33.3(14)$ and $z_\perp=0.97(15)$ as expected for the 2DHA. This result is also supported by the intensity ratio $I_\perp/I_\parallel$ shown in FIG.~\ref{Fig:Rb2MnF4.Gamma}(c), which approaches $1$ above $T_x$ as expected for the identical behavior of $\bm{M}_\parallel$ and $\bm{M}_\perp$ in the 2DHA.

\section{conclusions}

\begin{figure}
	\includegraphics[width=\columnwidth]{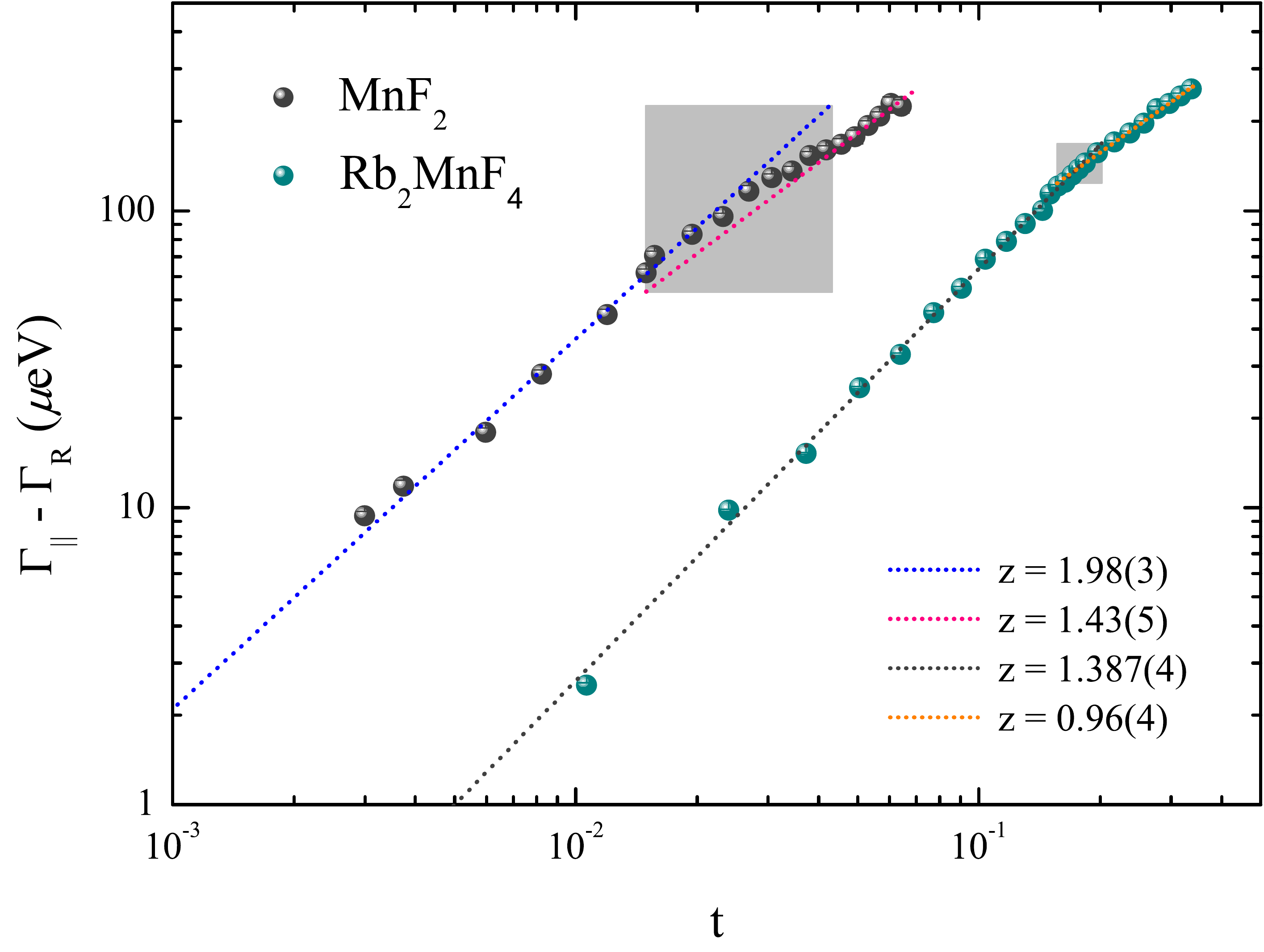}
	\caption{Scaling plot of the linewidth of longitudinal spin fluctuations in MnF$_2$ and Rb$_2$MnF$_4$. The residual linewidth at $T_N$ was subtracted from
the data.}
	\label{fig:Gamma_perp}
\end{figure}

FIG.~\ref{fig:Gamma_perp} summarizes the salient results of our study of the dynamical critical behavior of two canonical weakly anisotropic $S=5/2$ antiferromagnets with 3D and 2D spin coupling, respectively. Both compounds show a crossover in the scaling behavior resulting from the small uniaxial anisotropy induced by dipolar interactions. The dynamic critical exponent in MnF$_2$ changes from $z_\parallel=1.43(5)$ at high $T$, consistent with 3D Heisenberg scaling, to $z_\parallel=1.98(3)$ corresponding to a 3D Ising model close to $T_N$. This crossover occurs around $T_x=1.03T_N$, consistent with predictions in the literature. \cite{riedel1970,pfeuty1974} The previous contradictory experimental results for the longitudinal fluctuations, with $z_\parallel$ ranging from $1.6$ to $2.3$, are mainly due to the insufficient energy resolution of conventional triple-axis spectroscopy. The transverse linewidths $\Gamma_\perp$ are consistent with the predicted value $z_\perp=0$ around $T_x$, but $\Gamma_\perp$ decreases significantly upon cooling towards $T_N$. This behavior was also observed in earlier TAS experiments.

The dynamical critical exponent $z_\parallel$ measured in Rb$_2$MnF$_4$ changes around the crossover temperature $T_x = 1.18 T_N$ from $z_\parallel=0.96(4)$ for $T > T_x$, corresponding to the expected 2D Heisenberg scaling, to $z_\parallel = 1.387(4)$ for $T_N < T < T_x$. The latter value does not correspond to the expected $z = 1.75$ for the 2D Ising model. This scaling behavior probably results from the long-range nature of the dipolar forces, which influence the dynamic scaling in antiferromagnets by opening additional damping channels, while the static exponents remain unaffected. The transverse fluctuations show constant linewidths ($z_\perp = 0$) close to $T_N$ and are equal to the longitudinal fluctuations for $T \gg T_N$, where they show 2D Heisenberg scaling with $z_\perp = 0.97(15)$.

The high resolution three-axis spin-echo technique has thus provided detailed insight into the critical dynamics of antiferromagnets and helped resolve previous contradictory results. Our approach can straightforwardly be applied to a large class of questions on spin fluctuations and spin excitations, especially if a broad dynamic range with linewidths $<1\,\mu\text{eV}$ up to a few hundred $\mu\text{eV}$ has to be covered.

\acknowledgements

We thank the German Science Foundation (DFG) for financial support under grant No. SFB/TRR 80. We also thank Franz Tralmer for excellent technical support. The work at LBL was supported by the Director, Office of Science, Office of Basic Energy Sciences, Materials Science and Engineering Division, of the U.S. Department of Energy under Contract No. DE-AC02-05-CH11231 within the Quantum Materials Program (KC2202).

\bibliography{classical_AFs_2_V2_BK_Jason}
\end{document}